\documentstyle[12pt]{article}
\newcommand\bea{\begin{eqnarray}}
\newcommand\eea{\end{eqnarray}}
\begin{document}
\setlength{\baselineskip}{24pt}
\begin{center}{
{\Large 
{{\bf Uncertainty, entropy and decoherence  \\
of the damped harmonic oscillator \\
in the Lindblad theory of open quantum systems  }}
}\\
\vskip 1truecm
A. Isar$^{\dagger}$\\
{\it Department of Theoretical Physics, Institute of Atomic Physics \\
Bucharest, Magurele, POB MG-6, Romania }\\
}
\end{center}
\vskip 1truecm

\begin{abstract}

In the framework of the Lindblad theory for open quantum systems,
expressions for the density operator,
von Neumann entropy and effective
temperature of the damped harmonic oscillator
are obtained. The entropy for a state characterized by a Wigner
distribution function which is Gaussian in form is found to depend only on 
the variance of the distribution function.
We give a series of inequalities, relating uncertainty to von Neumann entropy
and linear
entropy. We analyze the conditions for purity of states and show that for a 
special choice of the diffusion coefficients, the correlated coherent states  
(squeezed coherent states) are 
the only states which remain pure all the time
during the evolution of the considered system.
These states are also the most stable under evolution in the presence of
the environment and play an important role in the description of
environment induced decoherence.

\vskip 0.5truecm
\end{abstract}


$^{\dagger}$ e-mail address: isar@theor1.ifa.ro

\newpage

\section{Introduction}

In the last years a large interest has arisen
to study open systems, especially in quantum
optics \cite{Car}, quantum measurement theory \cite{Zur2}
and in connection with decoherence and quantum to classical transition
\cite{GeH,JoZ,Zeh,gisp,omn}.
The consistent description of open quantum systems was investigated by various
authors \cite{h,d,s,de,l}  (for a recent review see Ref. \cite{rev}).
It is commonly understood \cite{de,d1} that dissipation
in an open system results from microscopic reversible interactions between
the observable system and the environment. 
Because dissipative processes imply irreversibility and,
therefore, a preferred direction in time, it is generally thought that quantum
dynamical semigroups are the basic tools to introduce dissipation in quantum 
mechanics.
The most general form of the generators of such semigroups was given
by Lindblad \cite{l1}, under the assumption that the evolution of the system
is Markovian.
This assumption is not generally valid, but it is considered
a good approximation for various interesting physical
models (for example, the quantum Brownian motion in a high
temperature reservoir).
In the Lindblad theory the
density operator properties (Hermiticity, unit trace and
positivity) are preserved by the master equation.
 
Different master equations for the evolution of the density operator
have been used to describe
decoherence and quantum to
classical transition
\cite{Zeh,gisp,omn,HPZ,paz1,PaZ,UnZ,paz2,Zur}.
In particular models, the density
operator becomes diagonal in some basis,
indicating that interference between the states in
that basis is suppressed -- this is the process of decoherence
of density operators.
In this case the dynamical variables
corresponding to the diagonalizing basis are expected to
evolve approximately classically.
A different theoretical approach to the
problem of quantum to classical transition is the decoherent
histories approach \cite{GeH,Gri,Omn,DoH}.

The Lindblad formalism has been studied for the case of the damped
harmonic oscillator \cite{rev,l2,ss,i3} and applied to various physical 
phenomena, 
for instance, the damping of collective modes in deep inelastic collisions in 
nuclear physics \cite{i1}.
In
\cite{g} a 
family of master equations, constructed in the form of Lindblad 
generators, was proposed for local Ohmic quantum dissipation.
In \cite{i2} the Lindblad master
equation for the harmonic oscillator
was transformed into Fokker-Planck equations for quasiprobability distributions
and a comparative study was made for the Glauber $P$, antinormal ordering $Q$ 
and Wigner $W$ representations. In \cite{wig,ent} the density matrix for
the coherent
state representation and the Wigner distribution function subject to different
types of initial conditions were obtained for the damped harmonic oscillator.
A phase space representation for open quantum 
systems within the Lindblad theory was given in \cite{vlas}.

In the present paper we are also concerned
with the observable system of a harmonic oscillator which interacts with
an environment. 
In Sec. 2 we give the description of the Lindblad model for open quantum
systems and write
the master equation for the density operator of the damped harmonic 
oscillator.
Generally the master equation gains considerably in clarity if it is 
represented in terms 
of the Wigner distribution function which satisfies the Fokker-Planck 
equation. 
In Sec. 3 we transform the
master equation into the Fokker-Planck equation by means of the 
well-known methods 
\cite{de,h1,lo,gr,h2,ga}
and analyze the evolution of the Wigner
function of the density operator. 
The Wigner function is shown to have a Gaussian form. Then 
we derive a closed form of the density operator
satisfying the master equation based on the Lindblad dynamics
and describe
the evolution of the considered system towards a final equilibrium state.
By using the explicit form of the density operator,
we calculate in Sec. 4 the von Neumann entropy and time dependent
temperature  of the quantum system in a state characterized
by a Wigner
distribution function which is Gaussian in form and analyze their
temporal evolution towards the equilibrium values.
Then we introduce the Wehrl entropy and present its basic
properties, including the Wehrl-Lieb inequality and 
another inequality which gives an interesting relationship between the
Wehrl entropy and von Neumann entropy. 
Following Halliwell and collaborators \cite{AndH,AnH},
we give two other simple inequalites, relating uncertainty to von Neumann 
entropy, and von Neumann entropy to linear
entropy. The dynamical behaviour of the Wehrl entropy is
compared to the von Neumann entropy and Shannon information
entropy (in some cases the two last entropies are equal to each other). 
The concept of the classical-like Wehrl entropy is a very informative
measure describing the time evolution of a quantum system. The Wehrl
entropy, first introduced as a classical entropy of a quantum state,
can give additional insights into the dynamics of the system, as compared to
other entropies.
We also discuss about the introduction of entropy production for
studying
the
stability of stationary states.
In Sec. 5 we analyze under what conditions the open system can be 
described by a quantum pure state
and show that for a special choice of the diffusion
coefficients, the correlated coherent states (squeezed coherent
states), taken as initial states,
remain pure for all time during the evolution of the system.   
In some simple models of the damped harmonic oscillator 
in the framework of quantum
statistical theory \cite{zel,hug}, it was shown that the pure 
Glauber coherent states remain as those during the evolution
and in all other cases, the oscillator immediately evolves into mixtures. 
In this respect  
we generalize this result,    
the results of other authors 
\cite{halzou}, obtained by using different methods as well as our previous
result of Ref. \cite{i3}.
In Sec. 6 we use the linear entropy 
for the description of the environment induced 
quantum decoherence phenomenon. We state that the 
correlated coherent states are the most stable under evolution 
in the presence of the environment and make the connection with the 
work done in this field by other authors \cite{paz1,paz2,AndH,AnH,gal}.
Finally, a summary and concluding remarks are given in Sec. 7.

\section{Lindblad model for the damped quantum harmonic oscillator}

The simplest dynamics for an open system which describes an irreversible 
process is a
semigroup of transformations introducing a preferred direction in time
\cite{d,s,l1}.
In Lindblad's axiomatic formalism of introducing dissipation in quantum
mechanics, the usual von Neumann-Liouville equation ruling the time evolution
of closed quantum systems is replaced by the following Markovian master
equation for the density operator $\hat\rho (t)$ in the Schr\"odinger 
picture
\cite{l1}:
\bea{d\Phi_{t}(\hat\rho )\over dt}=L(\Phi_{t}(\hat\rho )). 
\label{lin1}\eea
Here, $\Phi_{t}$ denotes the dynamical semigroup describing the irreversible 
time evolution of the open system in the Schr\"odinger representation and $L$ 
the infinitesimal generator of the dynamical semigroup $\Phi_t$. 
The condition for the Lindblad theory is that the time scale
considered for the open system (subsystem) should be very long
compared to the relaxation time of the heat bath (external
system), but shorter than the recurrence time of the total system
assumed as a closed finite system \cite{haa,tal}. 
Using the 
structural theorem of Lindblad \cite{l1}, which gives the most general form of 
the 
bounded, completely dissipative Liouville operator $L$, we obtain the explicit 
form of the most general time-homogeneous quantum mechanical Markovian master 
equation:
\bea   {d\hat\rho (t)\over dt}=-{i\over\hbar}[\hat H,\hat\rho (t)]
+{1\over 2\hbar}\sum_{j}([\hat V_j
\hat\rho (t),\hat V_j^\dagger]+[\hat V_j,\hat\rho (t)\hat V_j^\dagger]).  
\label{gener} \eea 
Here $\hat H$ is the Hamiltonian of the system in the absence of environment.
The operators $\hat V_j,\hat V_j^\dagger$
are bounded operators on the Hilbert space $\cal H$ of the Hamiltonian
and they model the effect of the environment.
 
We mention that the Markovian master equations found in the 
literature are of this form after some rearrangement of terms, even for
unbounded Liouville operators. In this connection we assume that the general 
form of the master equation given by (\ref{gener}) is also valid for unbounded
Liouville operators.

We consider an open system consisting of a particle moving in a
quadratic (harmonic oscillator) potential, coupled to an environment
described by non-Hermitian Lindblad
operators in (\ref{gener}) which are a linear combination of position and
momentum operators.
We impose a simple condition on the operators $\hat H,\hat V_j,\hat V_j^
\dagger$ 
that they are functions of the basic observables $\hat q$ and $\hat p$ of the
one-dimensional quantum mechanical system (with $[\hat q,\hat p]=i\hbar
I,$ where $I$ is the identity operator on $\cal H$) of such kind 
that the obtained model is exactly solvable. A precise version of this last 
condition is that linear spaces spanned by first degree (respectively second 
degree) noncommutative polynomials in $\hat q$ and $\hat p $ are invariant to 
the action 
of the completely dissipative mapping $L$. This condition implies \cite{l2} 
that $\hat V_j$ 
are at most first degree polynomials in $\hat q$ and $\hat p $ and $\hat H$ is 
at most a second degree polynomial in $\hat q$ and $\hat p $.
Because in the linear space of the first degree polynomials in $\hat q$ and
$\hat p$
the operators $\hat q$ and $\hat p$ give a basis,
there exist only two ${\bf C}$-linear
independent operators $\hat V_{1},\hat V_{2}$ which can be written in the form 
\bea \hat V_j=a_{j}\hat p+b_{j}\hat q,~j=1,2, \label{oper}\eea
with $a_{j},b_{j}$ complex numbers \cite{l2}. The constant term is omitted 
because its contribution to the generator $L$ is equivalent to terms in 
$\hat H$ 
linear in $\hat q$ and $\hat p$ which for simplicity are assumed to be zero.
Then the harmonic
oscillator Hamiltonian $\hat H$ is chosen of the general form 
\bea   \hat H=\hat H_{0}+{\mu \over 2}(\hat q\hat p+\hat p\hat q),
~~~\hat H_{0}={1\over 2m}
\hat p^2+{m\omega^2\over 2}\hat q^2.   \label{ham}  \eea 
With these choices the Markovian master equation can be written \cite{rev,ss}:
\bea   {d\hat\rho  \over dt}=-{i\over \hbar}[\hat H_{0},\hat\rho ]-
{i\over 2\hbar}(\lambda +
\mu)[\hat q,\hat\rho  \hat p+\hat p\hat\rho ]+{i\over 2\hbar}(\lambda -\mu)
[\hat p,\hat\rho  
\hat q+\hat q\hat\rho ]  \nonumber\\ 
  -{D_{pp}\over {\hbar}^2}[\hat q,[\hat q,\hat\rho ]]-{D_{qq}\over {\hbar}^2}
[\hat p,[\hat p,\hat\rho ]]+{D_{pq}\over {\hbar}^2}([\hat q,[\hat p,\hat\rho ]]
+[\hat p,
[\hat q,\hat\rho ]]).   \label{mast}   \eea 
Here we have used the notations:
\bea D_{qq}={\hbar\over 2}\sum_{j=1,2}{\vert a_{j}\vert}^2,
~~D_{pp}={\hbar\over 2}\sum_{j=1,2}{\vert b_{j}\vert}^2, \nonumber\\
~~D_{pq}=D_{qp}=-{\hbar\over 2}{\rm Re}\sum_{j=1,2}a_{j}^*b_{j},
~~\lambda=-{\rm Im}\sum_{j=1,2}a_{j}^*b_{j}, \label{coef}\eea
where $D_{qq},D_{pp}$ and $D_{pq}$ are the diffusion coefficients and $\lambda$
is the friction constant. They satisfy the following
fundamental constraints
\cite{rev,ss}:
\bea   {\rm i})~D_{pp}>0,~~{\rm ii})~D_{qq}>0,~~{\rm iii})~D_{pp}D_{qq}-D_{pq}
^2\ge{{\lambda}^2{\hbar}^2\over 4}.  \label{ineq}  \eea 
The semigroup method is valid for the weak coupling 
regime, with the damping
$\lambda$ typically obeying the inequality $\lambda\ll\omega_0,$ where
$\omega_0$ is the lowest frequency typical of reversible motions. 

In the particular case when the asymptotic state is a Gibbs state
\bea   \hat\rho _G(\infty)=e^{-{\hat H_0\over kT}}/
{\rm Tr}e^{-{\hat H_0\over kT}},  \label{gib}  \eea 
these coefficients reduce to
\bea   D_{pp}={\lambda+\mu\over 2}\hbar m\omega\coth{\hbar\omega\over 2kT},
~~D_{qq}={\lambda-\mu\over 2}{\hbar\over m\omega}\coth{\hbar\omega\over 2kT}, 
~~D_{pq}=0,  \label{coegib}  \eea 
where $T$ is the temperature of the thermal bath and the
fundamental constraints are satisfied only if $\lambda>|\mu|.$

The necessary and sufficient condition for $L$ to be translationally
invariant is $\lambda=\mu$ \cite{rev,l2,ss}. 
Translation invariance means that $[p,L(\rho)]=L([p,\rho]).$
In the following general values
for $\lambda$ and $\mu$ will be considered.

In the literature, master equations of the type (\ref{mast}) are encountered
in concrete theoretical models for the description of different 
physical phenomena in quantum optics \cite{Car,a2,da,lz}, in
treatments of the 
damping of collective modes in deep inelastic collisions of heavy ions 
\cite{j} or in the quantum description of the
dissipation 
for the one-dimensional harmonic oscillator \cite{de,l,gr,h2}. A 
classification of these equations, whether they satisfy or not the 
fundamental constraints (\ref{ineq}), was given in \cite{i3}.
Density operators satisfying the Lindblad equation give statistical
predictions in full agreement with experiment in a wide variety
of situations.
Lindblad type equations are also frequently used in studies of
quantum decoherence \cite{JoZ,Zeh,paz1,Zur}. For example, in the
much-studied quantum Brownian model \cite{CaL,Dio7}, the master
equation is the Lindblad master equation with a single Lindblad operator
\bea
\hat V=(2D)^{-1/2}(\hat q+2{i\over \hbar}\gamma D\hat p),
\label{brop}\eea
with 
$D=\hbar^2/8m\gamma kT$ (where $\gamma$ is the dissipation and
$T$ is the temperature of the environment) and $\hat H=\hat H_s+
(\gamma/2)\{\hat q,\hat p\},$ where $\hat H_s$ is the distinguished subsystem
Hamiltonian in the absence of the environment.
The Lindblad master equation
does not, in fact, completely agree with the master equations
given in a number of papers on quantum Brownian motion.
In particular, the master equation given by Caldeira and Leggett
\cite{CaL}
does not contain the term $ [ \hat p, [ \hat p, \hat\rho ] ] $
and is known to violate the positivity of the density operator
on short time scales \cite{Amb}.
This difference is principial, but practically
Lindblad and Caldeira-Leggett equations are similar for
high temperature, when the Markovian
approximation is only valid
and the extra term is negligible,  since its coefficient is
proportional to $T^{-1}$ \cite{Dio7}.

In the following we denote by $\sigma_{AA}$ the dispersion of the operator
$\hat A$, i.e. $\sigma_{AA}=<\hat A^2>-<\hat A>^2,$
where $<\hat A>\equiv\sigma_A={\rm Tr}(\hat\rho  \hat A)$ and 
${\rm Tr}\hat\rho=1.$
By $\sigma_{AB}=1/2<\hat A\hat B+\hat B\hat A>-<\hat A><\hat B>$ we denote 
the correlation of operators $\hat A$ and $\hat B.$
 
From the master equation (\ref{mast}) we obtain the following equations of 
motion for the expectation values and variances of the coordinate and momentum
\cite{rev,ss}:
\bea{d\sigma_{q}(t)\over dt}=-(\lambda-\mu)\sigma_{q}(t)+{1\over m}\sigma_{p}
(t), \label{eqmo1}\eea
\bea{d\sigma_{p}(t)\over dt}=-m\omega^2\sigma_{q}(t)-(\lambda+\mu)\sigma_{p}
(t) \label{eqmo2}\eea
and
\bea{d\sigma_{qq}(t)\over dt}=-2(\lambda-\mu)\sigma_{qq}(t)+{2\over m}
\sigma_{pq}(t)+2D_{qq},\label{eqmo3}\eea
\bea{d\sigma_{pp}\over dt}=-2(\lambda+\mu)\sigma_{pp}(t)-2m\omega^2
\sigma_{pq}(t)
+2D_{pp}, \label{eqmo4}\eea
\bea{d\sigma_{pq}(t)\over dt}=-m\omega^2\sigma_{qq}(t)+{1\over m}\sigma_{pp}(t)
-2\lambda\sigma_{pq}(t)+2D_{pq}.\label{eqmo5}\eea
In the underdamped case $(\omega>\mu)$ considered in this paper, with
the notation $\Omega^2\equiv\omega^2-\mu^2$, we obtain \cite{rev,ss}:
\bea\sigma_q(t)=e^{-\lambda t}((\cos\Omega t+{\mu\over\Omega}\sin\Omega t)
\sigma_q(0)+{1\over m\Omega}\sin\Omega t\sigma_p(0)), \label{sol1}\eea
\bea\sigma_p(t)=e^{-\lambda t}(-{m\omega^2\over\Omega}\sin\Omega t\sigma_q(0)+
(\cos\Omega t-{\mu\over\Omega}\sin\Omega t)\sigma_p(0)) \label{sol2}\eea
and $\sigma_q(\infty)=\sigma_p(\infty)=0.$
It is convenient to consider the vectors
\bea X(t)=\left(\matrix{m\omega\sigma_{qq}(t)\cr
\sigma_{pp}(t)/m\omega\cr
\sigma_{pq}(t)\cr}\right)\eea
and 
\bea D=\left(\matrix{2m\omega D_{qq}\cr
2D_{pp}/m\omega\cr
2D_{pq}\cr}\right).\eea
With these notations the solutions for the variances can be written in the 
form \cite{rev,ss}:
\bea X(t)=(Te^{-Kt}T)(X(0)-X(\infty))+X(\infty),\label{sol3}\eea
where the matrices $T$ and $K$ are given by  
\bea T={1\over 2i\Omega}\left(\matrix{\mu+i\Omega&\mu-i\Omega&2\omega\cr
\mu-i\Omega&\mu+i\Omega&2\omega\cr
-\omega&-\omega&-2\mu\cr}\right),\label{matr1}\eea
\bea K=\left(\matrix{2(\lambda-i\Omega)&0&0\cr
0&2(\lambda+i\Omega)&0\cr
0&0&2\lambda\cr}\right)\label{matr2}\eea
and 
\bea X(\infty)=(TK^{-1}T)D.\label{xinf}\eea
The formula (\ref{xinf}) is remarkable because it gives a very simple 
connection 
between the asymptotic values $(t \to \infty)$ of $\sigma_{qq}(t),
\sigma_{pp}(t), \sigma_{pq}(t)$
and the diffusion coefficients $D_{qq},~D_{pp},~ $ $D_{pq}:$ 
\bea\sigma_{qq}(\infty)={1\over 2m^2\lambda(\lambda^2+\omega^2-\mu^2)}
(m^2(2\lambda(\lambda+\mu)+\omega^2)D_{qq}
+D_{pp}+2m(\lambda+\mu)D_{pq}), \label{varinfq}\eea
\bea\sigma_{pp}(\infty)={1\over 2\lambda(\lambda^2+\omega^2-\mu^2)}((m\omega)^2
\omega^2D_{qq}+(2\lambda(\lambda-\mu)+\omega^2)D_{pp}-2m\omega^2(\lambda-
\mu)D_{pq}),\label{varinfp}\eea
\bea\sigma_{pq}(\infty)={1\over 2m\lambda(\lambda^2+\omega^2-\mu^2)}(-(\lambda+
\mu)(m\omega)^2D_{qq}+(\lambda-\mu)D_{pp}+2m(\lambda^2-\mu^2)D_{pq}).
\label{varinfpq}\eea
These relations show that the asymptotic values $\sigma_{qq}(\infty),
\sigma_{pp}(\infty),\sigma_{pq}(\infty)$ do not depend on the initial values
$\sigma_{qq}(0),\sigma_{pp}(0),\sigma_{pq}(0)$. 
In the considered underdamped case  
we have
\bea Te^{-Kt}T=-{e^{-2\lambda t}\over 2\Omega^2}
\left(\matrix{b_{11}&b_{12}&b_{13}\cr
b_{21}&b_{22}&b_{23}\cr
b_{31}&b_{32}&b_{33}\cr}\right),\label{xvar}\eea
where $b_{ij},$ $i,j=1,2,3$ are time-dependent oscillating functions 
given by
(3.78) in \cite{ss}.

\section{Wigner distribution function and density operator}

One useful way to study the consequences of the master equation (\ref{mast})
for the density operator of the one-dimensional damped harmonic 
oscillator is to transform it into more familiar forms, such as the 
equations for the $c$-number quasiprobability distributions Glauber 
$P$, antinormal ordering $Q$ and Wigner $W$ associated with the 
density operator \cite{i2}. In this case the resulting differential 
equations of the Fokker-Planck type for the distribution functions can 
be solved by standard methods 
\cite{h1,lo,gr,h2,ga}
employed in quantum
optics and 
observables directly calculated as correlations of these distribution 
functions.

In \cite{i2,wig,vlas} we have transformed the master equation (\ref{mast}) 
for the density operator into the following Fokker-Planck equation 
satisfied by the Wigner 
distribution 
function $W(q,p,t):$ 
\bea   {\partial W\over\partial t}=
-{p\over m}{\partial W\over\partial q}
+m\omega^2 q{\partial W\over\partial p}
+(\lambda-\mu){\partial\over\partial q}(qW)
+(\lambda+\mu){\partial\over\partial p}(pW) \nonumber \\
+D_{qq}{\partial^2 W\over\partial q^2}    
+D_{pp}{\partial^2 W\over\partial p^2}    
+D_{pq}{\partial^2 W\over\partial p\partial q}.~~~~~~~~~~~~~~~~~~~   
\label{wigeq}\eea 
Since the drift coefficients are linear in the variables $q$ and 
$p$ and the diffusion coefficients are constant with respect to 
$q$ and $p,$ Eq. (\ref{wigeq}) describes an Ornstein-Uhlenbeck process
\cite{w}. Following the method developed by Wang and Uhlenbeck
\cite{w}, we 
solved in \cite{wig} this Fokker-Planck equation, subject to either the
wave packet type or the $\delta$-function type of initial conditions
in the underdamped case ($\omega>\mu$) of the harmonic oscillator.
One gets two-dimensional Gaussian distributions with 
different variances.
Wigner function allows us to compute
the moments of $q$ and $p$ at any time $t$ in terms of the initial moments.
By computing the long time limits of these moments, the form of the
long time limit of the Wigner function may be obtained, since it is
completely determined by its moments.

For an initial Gaussian Wigner function
the solution of Eq. (\ref{wigeq}) is 
\bea   W(q,p,t)={1\over 2\pi\sqrt{\sigma(t)}}
~~~~~~~~~~~~~~~~~~~~~~~~~~~~~~~~~ \nonumber \\
\times\exp\{-{1\over 2\sigma(t)}[\sigma_{pp}(t)(q-\sigma_q(t))^2+
\sigma_{qq}(t)(p-\sigma_p(t))^2-2\sigma_{pq}(t)(q-\sigma_q(t))(p-\sigma_p(t))]
\},\label{wig} \eea
where $\sigma(t)$ is the determinant of the dispersion (correlation)
matrix $M(t)$,
\bea\sigma(t)=\det M(t) =\sigma_{qq}(t)\sigma_{pp}(t)-\sigma_{pq}^2(t)
\label{det}\eea
and
\bea M(t)=\left(\matrix{\sigma_{qq}(t)&\sigma_{pq}(t)\cr
\sigma_{pq}(t)&\sigma_{pp}(t)\cr}\right). \label{sigma}\eea
We see that the initial Wigner function remains Gaussian and therefore the   
property of positivity is preserved in time.
When time $t\to\infty,$ $\sigma_q(t)$ and $\sigma_p(t)$ vanish, all
dependence on $\sigma_q(0)$ and $\sigma_p(0)$ drops out of the exponentials
in $W$ and
we obtain the steady state solution:
\bea   W_{\infty}(q,p)={1\over 2\pi\sqrt{\sigma(\infty)}}
\exp\{-{1\over 2\sigma(\infty)}[\sigma_{pp}(\infty)q^2+
\sigma_{qq}(\infty)p^2-2\sigma_{pq}(\infty)qp]
\},\label{wiginf} \eea
where
$\sigma(\infty)=
\sigma_{qq}(\infty)\sigma_{pp}(\infty)-\sigma^2_{pq}(\infty)$
and $\sigma_{qq}(\infty),\sigma_{pp}(\infty),\sigma_{pq}(\infty)$
are given by Eqs. (\ref{varinfq}) -- (\ref{varinfpq}).
All stationary solutions to the evolution
equations obtained in the long time limit
are possible as a result
of a balance between the wave packet spreading induced by the
Hamiltonian and the localizing effect of the Lindblad operators.

We obtain now the explicit form of the density operator
of the damped harmonic oscillator in the Lindblad theory by
using a technique analogous to those applied in the description 
of quantum relaxation \cite{a1,j1,lw,m}.
Namely, we apply,
like in \cite{a1,j1}, the relation 
$\hat\rho =2\pi\hbar {\bf\it N}\{W_s(q,p)\},$ where $W_s$ is the Wigner 
distribution
function in the form of standard rule of association and ${\bf\it N}$ is the 
normal ordering operator \cite{lo,wx} which acting on the function $W_s(q,p)$
moves $p$ to the right of $q.$ By the standard 
rule of association we mean the correspondence $p^mq^n\to \hat q^n\hat p^m$ 
between functions of two classical variables $(q,p)$ and functions of two 
quantum canonical operators $(\hat q,\hat p).$ 
The Wigner distribution function (\ref{wig}), which is in the form of the
Weyl rule of association \cite{hw}, can be transformed into the form of 
standard rule of association \cite{clm} by performing the operation:
\bea W_s(q,p)=\exp({1\over 2}i\hbar{\partial^2\over\partial p\partial q})
W(q,p).
\label{swig}\eea 
The normal ordering operator $N$ can be applied upon the Wigner function 
$W_s$ in 
Gaussian form by using McCoy's theorem \cite{lo,wx}.
Following Jang \cite{j1}, we obtained in \cite{ent} the
following expression of the density operator:
\bea  \hat\rho(t) ={\hbar\over \sqrt{\delta}}\exp\{{1\over 2}\ln{4\delta\over 
4\sigma(t)-
\hbar^2}-{1\over 2\hbar\sqrt{\sigma(t)}}\cosh^{-1}(1+{2\hbar^2
\over 4\sigma(t)-\hbar^2})\nonumber\\
  \times  [ \sigma_{pp}(t)(\hat q-\sigma_q(t))^2+\sigma_{qq}(t)(\hat p-
\sigma_p(t))^2-2\sigma_{pq}(t)
[(\hat q-\sigma_q(t))(\hat p-\sigma_p(t))-i{\hbar\over 2}] ] \},
\label{dens}      \eea 
where
\bea  \delta=\sigma_{qq}(t)\sigma_{pp}(t)-(\sigma_{pq}(t)-i{\hbar\over 2})^2.
\label{delta}\eea 
The density operator (\ref{dens}) has a Gaussian form, as expected from the
initial form of the Wigner distribution function. While the Wigner distribution
is expressed in terms of real variables $q$ and $p,$ the density operator is
a function of operators $\hat q$ and $\hat p.$ When time $t\to\infty,$
the density operator tends to
\bea \hat\rho (\infty)={2\hbar\over{\sqrt{4\sigma(\infty)-\hbar^2}}}
~~~~~~~~~~~~~~~~~~~~~~~~~~~~~~\nonumber\\
\times\exp\{-{1\over 
2\hbar\sqrt{\sigma(\infty)}}\ln{2\sqrt{\sigma(\infty)}+\hbar\over 2\sqrt
{\sigma(\infty)}-\hbar}
[\sigma_{pp}(\infty)\hat q^2+\sigma_{qq}(\infty)\hat p^2-\sigma_{pq}(\infty)
(\hat q\hat p+\hat p\hat q)]\}. \label{densinf}   \eea 
In the particular case (\ref{coegib}),
\bea  \sigma_{qq}(\infty)={\hbar\over 2m\omega}\coth{\hbar\omega\over 2kT},~
\sigma_{pp}(\infty)={\hbar m\omega\over 2}\coth{\hbar\omega\over 2kT},~
 \sigma_{pq}(\infty)=0   \label{vargib} \eea 
and the asymptotic state is a Gibbs state (\ref{gib}):
\bea  \hat\rho _G(\infty)=2\sinh{\hbar\omega\over 2kT}\exp\{-{1\over kT}({1
\over 2m}\hat p^2+{m\omega^2\over 2}\hat q^2)\}.\label{densgib}\eea   

The Wigner function can be expressed as the Fourier transform of the 
off-diagonal matrix elements of the density operator in the coordinate
representation \cite{h5}:  
\bea W(q,p)={1\over \pi\hbar}\int dy<q-y|\hat\rho |q+y>e^{2ipy/\hbar}.
\label{fouri} \eea 
Then $<x|\hat\rho |y>$ can be obtained from the inverse Fourier
transform of the Wigner function:
\bea <x|\hat\rho |y>=\int dp \exp({i\over \hbar}p(x-y))W(p,{x+y\over 2}),
\label{fourinv}\eea
namely
\bea <x|\hat\rho(t)|y>=({1\over 2\pi\sigma_{qq}(t)})^{1\over 2} 
\exp[-{1\over 2\sigma_{qq}(t)}({x+y\over 2}-\sigma_q(t))^2 ~~~~~~~~\nonumber \\
-{1\over 2\hbar^2}
(\sigma_{pp}(t)-{\sigma_{pq}(t)^2\over\sigma_{qq}(t)})(x-y)^2 
+{i\sigma_{pq}(t)\over\hbar\sigma_{qq}(t)}({x+y\over 2}-\sigma_q(t))(x-y)+
{i\over \hbar}\sigma_p(t)(x-y)].\label{ccd}\eea
In the long time limit $\sigma_q(t)=0, \sigma_p(t)=0$ and we have
\bea <x|\hat\rho(\infty)|y>=({1\over 2\pi\sigma_{qq}(\infty)})^{1\over 2}
~~~~~~~~~~~\nonumber \\ 
\times\exp[-{1\over 8\sigma_{qq}(\infty)}(x+y)^2
-{1\over 2\hbar^2}
(\sigma_{pp}(\infty)-{\sigma_{pq}(\infty)^2\over\sigma_{qq}(\infty)})(x-y)^2 
+{i\sigma_{pq}(\infty)\over 2\hbar\sigma_{qq}(\infty)}(x^2-y^2)].
\label{dinf}\eea

\section{Entropy and uncertainty}

The most natural measure of uncertainty of the quantum mechanical state
is the entropy.
Physically, entropy can be interpreted as a measure
of the lack of knowledge (disorder) of the system.
The von Neumann entropy measures deviations from pure state
behaviour. For an isolated system the entropy is time
independent due to the unitarity of the evolution operator. For
open systems such as the damped harmonic oscillator, 
the evolution is not unitary and the entropy becomes time
dependent. Denoting by $ \hat\rho (t)$ the density operator of the damped 
harmonic 
oscillator in the Schr\"odinger
picture, the von Neumann entropy  $S(t) $ is given by the expectation value
of the logarithmic operator $\ln\hat\rho$ (we put Boltzmann's
constant $k=1):$
\bea  S(t)=-<\ln\hat\rho(t) >=-{\rm Tr}(\hat\rho(t)\ln\hat\rho(t) ).
\label{neuent1}\eea
Accordingly, the calculation of the entropy reduces to the
problem of finding the explicit form of the density operator.
Using Eq. (\ref{dens}) 
we obtained in \cite{ent} the following expression of the von Neumann
entropy:
\bea  S(t)=(\nu+1)\ln(\nu+1)-\nu\ln\nu, \label{neuent2}   \eea
where we denote $\hbar\nu=\sqrt{\sigma(t)}-\hbar/2.$
It is worth noting that the entropy depends only upon the variance of the 
Wigner distribution. 
When time $t\to\infty,$  $\nu$ tends to 
$s\equiv{\sqrt{\sigma(\infty)}/\hbar}-1/2$
and the entropy relaxes to its equilibrium value $S(\infty)=(s+1)\ln(s+1)-s
\ln s.$
The expression (\ref{neuent2}) is analogous to those previously obtained 
\cite{lw,rh}
in the theory of quantum oscillator relaxation
and for the description of a system of collective RPA phonons \cite{j1}.
It should also be noted that the expression (\ref{neuent2}) has the same form 
as the
entropy of a system of harmonic oscillators in thermal equilibrium. In the 
later case $\nu$ represents the average of the number 
operator \cite{a1}.
Although the expression (\ref{neuent2}) for the entropy has a well-known
form, the function $\nu$ induces a specific behaviour of the entropy.  
From the expression of the variances (\ref{sol3}), (\ref{xvar}),
which appear in (\ref{det}), we can see that the time
dependence of the entropy is given by the damping factors $\exp(-4
\lambda t),~ \exp(-2\lambda t) $ and the oscillating function $\exp 
2i\Omega t.$ The complex
oscillating factor $\exp 2i\Omega t$ reduces to a function of the 
frequency $\omega,$ 
namely $\exp 2i\omega t,$ for $\mu\to 0$ or if $\mu/\Omega\ll 1$ (i.e.
the frequency $\omega$ is very large as compared to $\mu).$ 

The von Neumann entropy gives zero for all pure states $\hat\rho =\hat\rho ^2,$
so it measures the purity of quantum states, being different
from zero (in fact positive) only
for mixed states. This entropy does not differentiate between various
pure states.
We have pure states for $\sigma=\hbar^2/4,$ which is just the case when
we have equality in the generalized uncertainty relation
$\sigma \ \ge \ \hbar^2/4$ \cite{dodkur} (see Sec. 5).
By calculating
\bea {dS\over d\nu}=\ln(1+{1\over\nu}),\label{derent}\eea
we see that $S$ is increasing when $\nu$ increases. But in general the entropy
is not a monotonic function of time, because we cannot decide a priori about
the sign of $dS/dt.$ In fact, in the underdamped case considered in this paper,
the entropy oscillates and only when $t\to\infty$ it relaxes to its equilibrium
value, which is independent of the initial state of the system.
If we require that $dS(t)/dt\ge 0,$ then the environment
operators $\hat V_j$ must be Hermitian. This assertion was proved 
in Ref. \cite{raj}
by
a direct calculation and the use of the inequality 
$(x-y)\log(x/y)\ge 0$ for positive $x,y$.

In the case of a thermal bath (\ref{gib}), a time dependent effective
temperature $T_e$ can be defined \cite{j1,lw},
by noticing that when $t\to\infty,$ 
$\nu$ tends, according to (2.24) in Ref. \cite{wig}, to the average thermal
phonon number $<n>=(\exp(\hbar\omega/kT)-1)^{-1}.$ Thus $\nu$ can be
considered as giving the time evolution of the thermal phonon number,
so that we can put in this case
\bea  (\exp{\hbar\omega\over kT_e}-1)^{-1}=\nu. \label{thernu}\eea 
From (\ref{thernu}) the effective temperature $T_e$ can be expressed as
\bea  T_e(t)={\hbar\omega\over k[\ln(\nu+1)-\ln\nu]}.\label{temp}\eea 
Accordingly, we can say that at
time $t$ the system is in thermal equilibrium at temperature $T_e.$ 
In terms of the effective temperature, the von Neumann entropy takes the form
\cite{ent}
\bea S(t)={\hbar\omega\over T_e(\exp{\displaystyle {\hbar\omega\over kT_e}}
-1)}-k\ln[1-\exp(-{\hbar\omega\over kT_e})].\label{tempent}\eea
As $t$ increases, the
effective temperature approaches thermal equilibrium with the bath, $T_e\to T.$

The von Neumann entropy of the density operator
is often connected with
uncertainty, decoherence and correlations of the distinguished
system with its environment \cite{Zur2,JoZ,paz1,PaZ,paz2}.
Zurek, Paz and Habib,
for example, looked for classes of initial
states which generate the least amount of entropy at time $t$. They
regarded such states as the most stable under evolution in
the presence of an environment. They argued that these states
are coherent states, at least approximately. The coherent states
are known to be as close as possible to classical states:
quantum fluctuations are minimal in these states and equal to
those of the vacuum.
Anastopoulos and Halliwell have shown that it is really the correlated
coherent states, rather than the ordinary coherent states
which are the most stable \cite{AnH}.
One of the reasons for looking at the von Neumann entropy is that it is
constant for unitary evolution, thus for open systems such as that
considered here, it is mainly a measure of environmentally
induced effects (see Sec. 6).

Entropic uncertainty relations have been obtained by
Bialynicki-Birula in Refs. \cite{bia1,bia2}. These relations
express restrictions imposed by quantum theory on probability
distributions of canonically conjugate variables in terms of
corresponding information entropies \cite{bia1}. Entropic
uncertainty relations derived from phase space quasiprobability
distributions and connection to the decay of quantum coherences
have been discussed in Refs. \cite{buz4}. Further discussion on
entropic uncertainty relations associated with the phase-number
uncertainties can be found in Ref. \cite{orl}. 

Following \cite{AndH,AnH}, we shall give the connection
between uncertainty and the von Neumann entropy,
by considering the phase space quasiprobability $Q_H$ distribution
\bea Q_H(p,q)=<\alpha|\hat\rho |\alpha>,
\label{QH} \eea
where 
\bea <x|\alpha> =<x|p,q>={1\over (2\pi s_{qq})^{1/4}}
\exp(-{(x-q)^2\over 4 s_{qq}}+{i\over \hbar}px) \label{stcs}\eea
are the standard coherent states,
$s_{qq}$ and $s_{pp}$ are the variances of the coherent 
states and
$ s_{qq} s_{pp} = \hbar^2/4 $.
It is normalized according to
\bea
\int { dp dq \over 2 \pi \hbar} Q_H(p,q)=1.
\label{norm}\eea
Between the two quasiprobability distributions $Q_H$ and Wigner
$W$ there is the following relation \cite{AnH}:
\bea
Q_H(p,q) = 2 \int dp' dq' \ \exp
\left( - { (p-p')^2 \over 2 s_{pp} } - { (q-q')^2
\over 2 s_{qq} } \right)  W (q',p').
\label{hus}\eea
Eq. (\ref{hus}) is the Husimi distribution \cite{Hus}.
In contrast to the other two quasiprobability distributions Glauber $P$
and Wigner $W$, the $Q_H$ representation is always a positive and
well-behaved function \cite{h5,Hal1}.
The distribution $Q_H(p,q)$ is therefore a Wigner function,
Gaussian smeared over an $\hbar$-sized region of phase space.

An information-theoretic
measure of the uncertainty in phase space contained in the
distribution (\ref{hus}),
is given by the Shannon information of the $Q_H$ distribution:
\bea
I = - \int { dp dq \over 2 \pi \hbar} Q_H(p,q) \ln Q_H(p,q).\label{inf}
\eea
This is the so-called Wehrl entropy \cite{Weh}, defined in full analogy
to the classical entropy in phase space. Wehrl considered this
entropy as a classical analogue of the von Neumann entropy
and it was introduced as a classical entropy of a quantum state.
It gives additional insight into the system dynamics, as compared
to other entropies. 
$I$ is large for
spread out distributions and small for very concentrated ones.
The uncertainty principle implies that a genuine phase space
probability distribution in quantum mechanics cannot be
arbitrarily peaked about a point in phase space. The information
(\ref{inf}) should possess a lower bound and since the coherent
states are the states most concentrated in phase space, we
expect the lower bound to be the value of $I$ on a coherent
state \cite{AnH}.
The most important property of this classical-like entropy is
the following inequality, which gives a lower bound on the Shannon 
information:
\bea
I \  \ge  \ 1,\label{binf}
\eea
with equality if and only if the considered state $ \hat\rho  $
is a coherent state \cite{Lie}. The Wehrl entropy is a good measure of the
strength of the coherent component and it clearly distinguishes
coherent states, in other words it measures how close
a given state is to the coherent states or how much
coherence a given state has. Wehrl entropy is very sensitive
to the phase space dynamics (such as, e. g., spreading) of the $Q_H$
representation. It extracts from the $Q_H$ function essential
information about the investigated system. The Wehrl entropy cannot be
negative. This follows from the fact that $0\le Q_H(\alpha)\le 1$
and from the normalization condition (\ref{norm}). The $Q_H$ function can
never be so concentrated as to make $I$ negative. On the
contrary, classical distributions can be arbitrarily
concentrated in phase space and classical entropies can take on
negative values. They may even tend to $-\infty$ if the
distributions tend to $\delta$-functions.

The Shannon information can be used \cite{AnH} to find a relation
between the von Neumann entropy and the generalized uncertainty measure
\bea
\sigma \equiv \sigma_{qq}\sigma_{pp}-\sigma_{pq}^2. \label{uncme} 
\eea
The Shannon information satisfies the inequality
\bea
I \  \le  \ \ln \left( { e \over \hbar} (\det \sigma^{(Q)})^{1\over 2}
\right),\label{infin}
\eea
where $\sigma^{(Q)}$ is the $2 \times 2$ covariance matrix of the distribution
$Q_H(p,q)$ \cite{AnH,Cov}.
Equality holds if and only if $Q_H(p,q)$ is a Gaussian.
From (\ref{hus}) or using the results of Ref. \cite{i2}, one has
\bea
\det \sigma^{(Q_H)} =
\left(  \sigma_{qq} + s_{qq} \right) \left( \sigma_{pp} +
s_{pp} \right)
- \sigma_{pq}^2.\label{det1}
\eea
There is an even stronger relationship \cite{Weh}
\bea
I  \ \ge \ S, \label{infent}
\eea
which establishes a connection
between the Wehrl entropy and the von Neumann entropy of a given state.
Using the relations (\ref{infin}) -- (\ref{infent}),
one
obtains
\bea
\left(  \sigma_{qq} + s_{qq} \right) \left( \sigma_{pp} +
s_{pp} \right)
- \sigma_{pq}^2
 \ \ge \ \hbar^2 e^{2(S-1)}.\label{varent}
\eea
Since $s_{qq}$ is arbitrary and $s_{qq}s_{pp} =
\hbar^2/4 $, we may minimize the left-hand side over it and we obtain
the following connection between the uncertainty and entropy for a general
mixed state $\hat\rho $ \cite{AnH}:
\bea
( \sqrt{\sigma_{qq} \sigma_{pp}}  + { \hbar \over 2})^2 -\sigma_{pq}^2 \
\ge \
\hbar^2 e^{2(S-1)}.\label{uncent}
\eea
An analogous relation was obtained by Dodonov and Man'ko in \cite{dodman1}. 

In the regime where quantum fluctuations are more significant
than thermal ones, it is appropriate to use the lower bound
(\ref{binf}) rather than (\ref{infent}) (since $S$ goes to zero
if the state is pure) and this is formally achieved by setting
$S=1$ in (\ref{uncent}). One then deduces the usual uncertainty
principle from (\ref{uncent}) \cite{AnH}.

In the regime where thermal (or environmentally induced)
fluctuations are dominant, one would expect $\sigma_{qq}\sigma_{pp}\gg
\hbar^2/4$ and $S\gg 1$ and (\ref{uncent}) then gives 
\bea {\sqrt{\sigma_{qq}\sigma_{pp}}\over\hbar}\ge{{\cal A}\over\hbar}\ge e^S.
\label{uncent1}\eea
Here ${\cal A}=\sqrt{\sigma}$  is the Wigner function area
-- a measure of the phase space area in which the Gaussian
density matrix is localized and 
$\sqrt{\sigma_{qq}\sigma_{pp}}/\hbar $ 
is the number of phase space cells occupied by the state \cite{AnH}.
The von Neumann entropy of a Gaussian is given by
(\ref{neuent2}) and for large ${\cal A},$ (\ref{neuent2}) gives
\bea S\approx\ln{{\cal A}\over\hbar} \label{aent}\eea
and hence we have equality in (\ref{uncent1}) \cite{AnH}.

We can conclude that $I$ is a useful measure of both quantum and
thermal fluctuations. It possesses a lower bound expressing the
effect of quantum fluctuations and is closely connected to
entropy, which in turn is a measure of thermal fluctuations \cite{AndH}.

In some models the linear entropy is introduced as a measure of
purity of states:
\bea S_l ={\rm Tr}(\hat\rho-\hat\rho^2)=  1 - {\rm Tr} \hat\rho ^2.
\label{linent}
\eea
Since ${\rm Tr} \hat\rho ^2 \ \le \ 1,$ the linear entropy is positive
and it becomes zero if the state is pure. When the linear entropy
is increasing, the degree of purity is decreasing.
In \cite{AnH} the following relation is deduced between the linear
and von Neumann entropies:
\bea
S_l = 1 - {\rm Tr} \hat\rho ^2 \ \le \ 1 - e^{-S}.\label{lineq}
\eea
Equality in Eq. (\ref{lineq}) is reached for pure states, when $S = S_l = 0 $ 
and for very mixed states, when $S$ is very large and $S_l \approx  1$.

For analyzing the stability of stationary states, it could also be
interesting to introduce the entropy production
$\sigma_S(\hat\rho )$ \cite{PaS}, defined for dynamical semigroups as follows 
\cite{spo}:
\bea \sigma_S(\hat\rho )=-{d\over dt}S(\Phi_t(\hat\rho )|\hat\rho ^0)|_{t=0},
~~\hat\rho \in L_+^1(\cal H),\label{entprod}\eea
where
\bea S(\Phi_t(\hat\rho )|\hat\rho ^0)={\rm Tr}(\Phi_t(\hat\rho )
\ln \hat\rho ^0)+S(\Phi_t(\hat\rho )),
\eea
$\hat\rho ^0\in L_+^1(\cal H) $ is a $\Phi_t$-invariant state,
i.e. $\Phi_t\hat\rho ^0=\hat\rho ^0,$ for any $t\ge 0,$
$\Phi_t(\hat\rho)\equiv\hat\rho(t)$ and $L_+^1(\cal H) $
is the Banach space of trace class operators on $\cal H.$

\section{Purity of states}

Schr\" odinger \cite{schr} and Robertson \cite{rob} proved for any 
Hermitian operators
$\hat A$ and $\hat B$ and for pure quantum states
the following generalized  uncertainty relation :
\bea \sigma_{AA}\sigma_{BB}-\sigma_{AB}^2\ge {1\over 4}|<[\hat A,\hat B]>|^2.
\label{rob}\eea
For the particular case of the operators of the coordinate $\hat q$
and momentum $\hat p$ the uncertainty relation (\ref{rob}) takes the form
\bea \sigma_{qq}\sigma_{pp}-\sigma_{pq}^2 \ge{\hbar^2\over 4}.\label{genun}\eea
This result was generalized for arbitrary operators (in general non-Hermitian)
and for the most general case of mixed states in \cite{dodkur}.
The inequality (\ref{genun}) can also be represented in the following form:
\bea \sigma_{qq}\sigma_{pp}\ge{\hbar^2\over 4(1-r^2)},\label{dod}\eea
where 
\bea r={\sigma_{pq}\over\sqrt{\sigma_{qq}\sigma_{pp}}} \label{corcoe}\eea
is the correlation coefficient.
The equality in the relation (\ref{genun}) is realized for a special class of 
pure states, called correlated
coherent states \cite{dodkur} or squeezed coherent states,
which are represented by Gaussian wave packets 
in the coordinate representation.
These minimizing states, which generalize
the Glauber coherent states, are eigenstates of an operator of the form 
\cite{dodkur}:
\bea \hat a_{r,\eta}={1\over 2\eta}[1-{ir\over (1-r^2)^{1/2}}]\hat q+
i{\eta\over\hbar}\hat p, \label{eigv}\eea
with real parameters $r$ and $\eta,$ $|r|<1, \eta=\sqrt{\sigma_{qq}}.$ 
Their
normalized eigenfunctions, the correlated coherent states, have the form
\cite{dodkur}:
\bea \Psi(x)={1\over (2\pi\eta^2)^{1/4}}\exp\{-{x^2\over 4\eta^2}[1-
{ir\over (1-r^2)^{1/2}}]+{\alpha x\over\eta}-{1\over 2}(\alpha^2+
|\alpha|^2)\}, \label{eigf} \eea 
with $\alpha$ a complex number.
If we set $r=0$ and $\eta=(\hbar/2m\omega)^{1/2},$ where $m$ and $\omega$ are
the mass and respectively the frequency of the harmonic oscilator,
the states (\ref{eigf}) become the usual
Glauber coherent states. In Wigner representation, the states (\ref{eigf}) 
have   the form \cite{dodkur}:
\bea W_{\alpha,r,\eta}(q,p)=
{1\over\pi\hbar}\exp[-{2\eta^2\over\hbar^2}
(p-\sigma_p)^2-{(q-\sigma_q)^2\over 2\eta^2(1-r^2)}+{2r\over \hbar(1-r^2)^
{1/2}}(q-\sigma_q)(p-\sigma_p)]. \label{corwig}\eea
This is the classical normal distribution to give dispersion 
\bea \sigma_{qq}=\eta^2,~~ \sigma_{pp}={\hbar^2\over 4\eta^2(1-r^2)},~~
\sigma_{pq}={\hbar r\over 2({1-r^2})^{1/2}} \label{disp} \eea
and the correlation coefficient $r.$
The Gaussian distribution (\ref{corwig}) is the only positive Wigner 
distribution for a pure state \cite{huds}. All other Wigner functions
that describe pure states necessarily take on negative values for some
values of $q,p.$

In the case of the relation (\ref{rob}) the equality is generally obtained 
only for pure states \cite{dodkur}. For any density matrix in the coordinate 
representation (normalized to unity)
the following relation must be fulfilled:
\bea \gamma={\rm Tr}\hat\rho ^2\leq 1.
\label{ropur}\eea
The quantity $\gamma$ characterizes the degree of purity of the state.  
For pure states $\gamma=1,$ for highly mixed states $\gamma\ll 1$
and for weekly mixed states $1-\gamma\ll 1.$ 
In the language of the Wigner function the condition (\ref{ropur})
has the form:
\bea \gamma=2\pi\hbar\int W^2(q,p)dqdp\leq 1.
\label{wigpur}\eea

Let us consider the most general mixed squeezed states described by the 
Wigner function of the generic Gaussian form with five real parameters:
\bea   W(q,p)={1\over 2\pi\sqrt{\sigma}}\exp\{-{1\over 2\sigma}[\sigma_{pp}
(q-\sigma_q)^2+\sigma_{qq}(p-\sigma_p)^2-2\sigma_{pq}(q-\sigma_q)(p-\sigma_p)]
\},  \label{gaus}  \eea 
where $\sigma$ is given by Eq. (\ref{uncme}).
The Gaussian Wigner functions of this form correspond to the so-called 
quasi-free states on the $C^*$-algebra of the canonical commutation relations,
which is the most natural framework for a unified treatment of quantum and 
thermal fluctuations \cite{scut}.
For Gaussian states of the form (\ref{gaus}) the coefficient of purity 
$\gamma$ is given by
\bea \gamma={\hbar\over 2\sqrt{\sigma}}. \label{pursig}\eea
Therefore, we reobtain from (\ref{ropur}) and (\ref{pursig}) that
$\sigma$ has to
satisfy the Schr\"odinger-Robertson uncertainty relation (\ref{genun})
\bea\sigma\geq {h^2\over 4}.\label{detsig}\eea
This inequality must be fulfilled actually for any states, not only Gaussian.
Any Gaussian pure state minimizes the relation (\ref{detsig}).
For $\sigma>\hbar^2/4$ the function (\ref{gaus}) corresponds to mixed quantum
states, while in the case of the equality $\sigma=\hbar^2/4$ it
takes the form (\ref{corwig}) corresponding 
to pure correlated coherent states (squeezed coherent states).

We have seen in the preceding section that the degree of purity of a state 
can also be characterized by other
quantities besides $\gamma.$ The most usual one is the quantum
(von Neumann) entropy.
For quantum pure states this entropy is
identically equal to zero.
We remind \cite{ent,a1} that for Gaussian states with the Wigner 
functions (\ref{gaus})
the entropy can be expressed through $\sigma$ only, in the form 
(\ref{neuent2}).

In the Lindblad model for the damped harmonic oscillator,
the relation (\ref{ineq}) is a necessary condition for the
generalized uncertainty inequality (\ref{genun}) to be fulfilled
\cite{rev,ss,i3}.
By using the fact that
the linear positive mapping 
defined by $\hat A\to {\rm Tr}(\hat\rho\hat A)$
is completely positive,  the
following inequality was obtained  in Refs. \cite{rev,ss}:
\bea D_{pp}\sigma_{qq}(t)+D_{qq}\sigma_{pp}(t)-2D_{pq}\sigma_{pq}(t)\ge
{\hbar^2\lambda\over 2}\label{has3}.\eea 
We have found in \cite{i3} that this inequality, which must be valid for all 
values of $t,$ is equivalent with the generalized
uncertainty inequality (\ref{genun}) at any time $t,$
\bea\sigma_{qq}(t)\sigma_{pp}(t)-\sigma_{pq}^2(t)\ge{\hbar^2
\over 4},\label{genun1}\eea
if the initial values $\sigma_{qq}(0),\sigma_
{pp}(0)$ and $\sigma_{pq}(0)$ for $t=0$ satisfy this inequality.
If the initial state is the ground state of the harmonic oscillator, then
\bea\sigma_{qq}(0)={\hbar\over 2m\omega},~\sigma_{pp}(0)={m\hbar\omega\over 2},
~\sigma_{pq}(0)=0. \label{grovar1}\eea

By using the complete positivity property of the dynamical semigroup $\Phi_t,$
it was shown in Ref. \cite{ss} that the relation
\bea{\rm Tr}(\Phi_t(\hat\rho)\sum_j\hat V_j^\dagger\hat V_j)=\sum_j{\rm Tr}
(\Phi_t(\hat\rho)\hat V_j^\dagger){\rm Tr}(\Phi_t(\hat\rho)\hat V_j) 
\label{has1}\eea
represents the necessary and sufficient condition for $\hat\rho(t)=
\Phi_t(\hat\rho)$ to be a pure state for all times $t\ge 0.$
This equality is a generalization of the pure state condition 
\cite{dek,has1,has2}
to all Markovian master equations (\ref{gener}).
If $\hat\rho^2(t)=\hat\rho(t)$ for all $t\ge 0,$ there exists a
wave function
$\psi\in{\cal H}$ which satisfies the nonlinear Schr\"odinger type equation
\bea i\hbar{\partial \psi(t)\over \partial t}=\hat H'\psi(t), \label{scheq}
\eea
with the non-Hermitian Hamiltonian \cite{ss}
\bea \hat H'=\hat H+i\sum_j<\psi(t),\hat V_j^\dagger\psi(t)>\hat V_j-{i\over 2}
<\psi(t),\sum_j\hat V_j^\dagger\hat V_j\psi(t)>-{i\over 2}\sum_j
\hat V_j^\dagger\hat V_j. \label{ham1} \eea
For environment operators $\hat V_j$ of the form (\ref{oper}), 
the pure state condition (\ref{has1}) takes the following form
\cite{ss}, corresponding to equality in the relation (\ref{has3}):
\bea D_{pp}\sigma_{qq}(t)+D_{qq}\sigma_{pp}(t)-2D_{pq}\sigma_{pq}(t)=
{\hbar^2\lambda\over 2}\label{has2}\eea
and the Hamiltonian (\ref{ham1}) becomes
\bea \hat H'=\hat H+\lambda(\sigma_p(t)\hat q-
\sigma_q(t)\hat p)+{i\over\hbar}[\lambda\hbar^2-D_{pp}((\hat q-
\sigma_q(t))^2+
\sigma_{qq}(t))
-D_{qq}((\hat p-
\sigma_p(t))^2\nonumber \\
+\sigma_{pp}(t))
+D_{pq}((\hat q-
\sigma_q(t))(\hat p-
\sigma_p(t))+(\hat p-
\sigma_p(t))(\hat q-
\sigma_q(t))+2
\sigma_{pq}(t))]. ~~~~~~~\label{ham2}\eea

We will find the Gaussian states which remain pure during the 
evolution of the system for all times $t.$ We start by considering the
pure state condition (\ref{has2}) and the generalized uncertainty relation
(\ref{genun}) which transforms into the following minimum uncertainty
equality for pure states:
\bea \sigma_{qq}(t)\sigma_{pp}(t)-\sigma_{pq}^2(t) ={\hbar^2\over 4}.
\label{pur}\eea
By eliminating $\sigma_{pp}$ between the equalities (\ref{has2}) and 
(\ref{pur}), like in Ref. \cite{deval}, we obtain:
\bea(\sigma_{qq}(t)-{D_{pq}\sigma_{pq}(t)+{1\over 4}\hbar^2\lambda\over
D_{pp}})^2+{D_{pp}D_{qq}-D_{pq}^2\over D_{pp}^2}[(\sigma_{pq}(t)-{{1\over 4}
\hbar^2\lambda D_{pq}\over D_{pp}D_{qq}-D_{pq}^2})^2 \nonumber\\
+{1\over 4}\hbar^2{D_{pp}D_{qq}-D_{pq}^2-{1\over 4}\hbar^2\lambda^2\over
(D_{pp}D_{qq}-D_{pq}^2)^2}D_{pp}D_{qq}]=0.~~~~~~~~~~~~~~~~~~ \label{dek}\eea
Since the opening coefficients satisfy the inequality (\ref{ineq}),
we obtain from Eq. (\ref{dek})
the following relations
which have to be fulfilled at any moment of time:
\bea D_{pp}D_{qq}-D_{pq}^2={\hbar^2\lambda^2\over 4}, \label{coe1}\eea
\bea D_{pp}\sigma_{qq}(t)-D_{pq}\sigma_{pq}(t)-{\hbar^2\lambda\over 4}=0,
\label{coe2}\eea
\bea \sigma_{pq}(t)(D_{pp}D_{qq}-D_{pq}^2)-{\hbar^2\lambda\over 4}D_{pq}=0.
\label{coe3}\eea
From the relations (\ref{pur}) and (\ref{coe1}) -- (\ref{coe3}) it
follows that the pure states remain pure for all times only if
their variances are constant in time and have the form:
\bea \sigma_{qq}(t)={D_{qq}\over\lambda},~~
\sigma_{pp}(t)={D_{pp}\over\lambda},~~
\sigma_{pq}(t)={D_{pq}\over\lambda}. \label{coe4}\eea
If these relations are fulfilled, then the inequalities
(\ref{has3}) and (\ref{genun1}) are both equivalent to (\ref{ineq}),
including also the corresponding equalities (\ref{has2}), (\ref{pur}) and
(\ref{coe1}). 
From Eq. (\ref{sol3}) it follows that the variances remain constant
and do not depend on time only if $X(0)=X(\infty),$ that means 
$\sigma_{qq}(0)=\sigma_{qq}(\infty),$
$\sigma_{pp}(0)=\sigma_{pp}(\infty),$$\sigma_{pq}(0)=\sigma_{pq}(\infty).$
Using the asymptotic values (\ref{varinfq}) -- (\ref{varinfpq}) of the 
variances and the
relations (\ref{coe4}), we obtain
the following expressions of the diffusion coefficients which
assure that the initial pure states remain pure for any $t:$
\bea D_{qq}={\hbar\lambda\over 2m\Omega},~~
D_{pp}={\hbar\lambda m\omega^2 \over 2\Omega},~~
D_{pq}=-{\hbar\lambda\mu\over 2\Omega}. \label{coepur}\eea
Formulas (\ref{coepur}) are generalized Einstein 
relations and represent examples of quantum 
fluctuation-dissipation relations, connecting the diffusion with both 
Planck's constant and damping constant.
With the coefficients (\ref{coepur}), the variances (\ref{varinfq}) -- 
(\ref{varinfpq}) 
become:
\bea \sigma_{qq}={\hbar\over 2m\Omega},~~
\sigma_{pp}={\hbar m\omega^2 \over 2\Omega},~~
\sigma_{pq}=-{\hbar\mu\over 2\Omega}. \label{varpur}\eea
Therefore, the quantity $\sigma$ (see Eqs. (\ref{det}), (\ref{uncme})) is equal
to its 
minimum possible
value $\hbar^2/4,$ according to the generalized uncertainty relation 
(\ref{detsig}). Then the corresponding state described by a Gaussian Wigner
function is a pure quantum state, namely a correlated coherent state
(squeezed coherent state) with the correlation coefficient 
(\ref{corcoe}) $r=-\mu/\omega.$
Given $\sigma_{qq},$ $\sigma_{pp}$ and $\sigma_{pq},$ there exists one and 
only one such a state minimizing the uncertainty $\sigma$ \cite{sud}. 
We remark that the minimization of the uncertainty $\sigma$ is equivalent,
by virtue of the relations (\ref{neuent2}), (\ref{derent})
to the minimization of the entropy $S$.
A particular case of our result (corresponding to $\lambda=\mu$ and 
$D_{pq}=0)$ was obtained by 
Halliwell and Zoupas by using the quantum state diffusion 
method \cite{halzou}. 
We consider here general
coefficients $\mu$ and
$\lambda$ and in this respect our expressions for the diffusion
coefficients and variances generalize also the ones  
obtained by Dekker and 
Valsakumar \cite{deval} and Dodonov and Man'ko \cite{dodman},
who used models where $\lambda=\mu$ was chosen.
If $\mu=0,$ we get from (\ref{coepur}) $D_{pq}=0.$ This case, which
was considered in Ref. \cite{i3} and where we have obtained a density operator 
describing a pure state
for any $t,$ is also a particular case of the present results which give the 
most general Gaussian pure state which remains pure for any $t.$
For $\mu=0$, the expressions (\ref{varpur}) become 
\bea \sigma_{qq}={\hbar\over 2m\omega},~~
\sigma_{pp}={\hbar m\omega\over 2},~~
\sigma_{pq}=0,   \label{grovar2}\eea
which are the values of variances (\ref{grovar1}) for the ground state of 
the harmonic oscillator
and the correlation coefficient (\ref{corcoe}) takes the value
$r=0,$ corresponding 
to the case of usual coherent states. 

The Lindblad equation or its equivalent Fokker-Planck equation for the Wigner
function with the diffusion coefficients (\ref{coepur}) can 
be used only in the underdamped case, when $\omega>\mu.$ 
Indeed, for the coefficients (\ref{coepur}) the fundamental constraint 
(\ref{ineq}) implies that $m^2(\omega^2-\mu^2)D_{qq}^2\ge\hbar^2\lambda^2/4,$
which is satisfied only if $\omega>\mu$.
It can be shown that
there exist diffusion coefficients which satisfy the condition
(\ref{coe1}) and make sense for $\omega<\mu,$ but in this overdamped case
we have always $\sigma>\hbar^2/4$ and the state of the oscillator cannot
be pure for any diffusion coefficients \cite{dodman}.   

The fluctuation energy of the open harmonic oscillator is
\bea E(t)={1\over 2m}\sigma_{pp}(t)+{1\over 2}m\omega^2\sigma_{qq}(t)+
\mu\sigma_{pq}(t). \label{ener}\eea
If the state remains pure in time, then the variances are given by
(\ref{coe4}) and the fluctuation energy is also constant 
in time and is given by
\bea E={1\over\lambda}({1\over 2m}D_{pp}+{1\over 2}m\omega^2D_{qq}+\mu D_{pq}).
\label{conen}\eea
Minimizing this expression with the condition (\ref{coe1}), we obtain
just the diffusion coefficients  (\ref{coepur}) and $E_{min}=\hbar\Omega/2.$
Therefore, the conservation of purity of state implies 
that the fluctuation energy of the system has all the time the minimum
possible value $E_{min}.$ 
If the asymptotic state is a Gibbs state (\ref{gib}), we see from
Eqs. (\ref{coegib}) and (\ref{coepur}) that the case when the 
diffusion coefficients satisfy the equality (\ref{coe1}) corresponds to a
zero temperature of the thermal bath and then the influence on the 
oscillator is minimal. In this limiting case $\mu=0$ and then 
$E_{min}=\hbar\omega/2,$ the correlation coefficient
(\ref{corcoe}) vanishes and therefore the correlated coherent
state (squeezed coherent state) becomes the usual coherent state.

If we choose the coefficients of the form (\ref{coepur}), then 
the equation for 
the density operator can be represented in the form (\ref{gener}) with only
one operator $\hat V,$ which up to a phase factor can be written in the form:
\bea \hat V=\sqrt{{2\over\hbar D_{qq}}}[({\lambda\hbar\over 2}-iD_{pq})\hat q+
iD_{qq}\hat p)], \label{oneop}\eea
with $[\hat V,\hat V^\dagger]=2\hbar\lambda.$   

The correlated coherent states 
(\ref{eigf}) with nonvanishing momentum average, can also be
written in the form:
\bea \Psi(x)=({1\over 2\pi\sigma_{qq}})^{1\over 4}\exp[-{1\over 4\sigma_{qq}}
(1-{2i\over\hbar}\sigma_{pq})(x-\sigma_q)^2+{i\over \hbar}\sigma_px]
\label{ccs}\eea
and the most general form of Gaussian density matrices compatible with the 
generalized uncertainty relation (\ref{genun}) is given by Eq. (\ref{ccd}).
For these matrices we can verify that 
${\rm Tr}\rho^2=\hbar/2\sqrt{\sigma}$
and they correspond to the correlated coherent states (\ref{ccs}) 
if $\sigma_{qq}, \sigma_{pp}$ and $\sigma_{pq}$ in Eq. (\ref{ccd}) satisfy the 
equality in Eq. (\ref{genun}).

Consider the harmonic oscillator initially in a correlated coherent state
(squeezed coherent state) of the form (\ref{ccs}), with the corresponding 
Wigner
function (\ref{corwig}). 
For an environment described by the diffusion coefficients (\ref{coepur}),
the solution for the Wigner function at time $t$ is given by (\ref{wig}),
where $\sigma_q(t)$ and $\sigma_p(t)$ are given respectively by
(\ref{sol1}) and (\ref{sol2}) and the variances by (\ref{varpur}).
Using Eq. (\ref{ccd}), we get for the 
density 
matrix the following time evolution: 
\bea <x|\hat\rho(t)|y>= ({m\Omega\over\pi \hbar})^{1\over 2}
\exp[-{m\Omega\over \hbar}({x+y\over 2}-\sigma_q(t))^2 ~~~~~~~~~~\nonumber \\
-{m\Omega\over 4\hbar}(x-y)^2
-{im\mu\over \hbar}({x+y\over 2}-\sigma_q(t))(x-y)+{i\over\hbar}\sigma_p(t) 
(x-y)]. \label{coord}\eea
In the long time limit $\sigma_q(t)=0, \sigma_p(t)=0$ and we have
\bea <x|\hat\rho(\infty)|y>= ({m\Omega\over\pi \hbar})^{1\over 2}
\exp\{-{m\over 2\hbar}[\Omega(x^2+y^2)+i\mu(x^2-y^2)]\}. \label{coorinf}\eea
The corresponding Wigner function (\ref{wiginf}) has the form
\bea   W_{\infty}(q,p)={1\over \pi\hbar} 
\exp[-{1\over \hbar\Omega}({1\over m}p^2+m\omega^2q^2+2\mu qp)].
\label{winf}  \eea 
We see that the time evolution of the initial correlated coherent state  
of the damped harmonic oscillator is given by a Gaussian density matrix with 
variances constant in time. According to known general results \cite{AndH,AnH},
the initial
Gaussian density matrix remains Gaussian centered around the classical path.
So, the correlated coherent state (squeezed coherent state) remains a 
correlated coherent 
state and $\sigma_q(t)$ and $\sigma_p(t)$ give the average time dependent
location of the system along its trajectory in phase space.   

\section{Decoherence and transition from quantum to classical} 

The environment induced quantum decoherence (loss of quantum coherence)
and the transition
from quantum to classical mechanics have recently been
investigated in widely different contexts 
\cite{GeH,JoZ,Zeh,gisp,omn,HPZ,paz1,PaZ,UnZ,paz2,Zur,Gri,Omn,DoH,AnH,gal}. 
The central idea of the quantum decoherence approach is that the
transition quantum -- classical is a dynamical effect within
quantum mechanics, keeping $\hbar\not=0.$
The characteristic feature of the decoherence process is that an
arbitrarily chosen generic initial quantum state 
will be affected
on a characteristic decoherence time scale and only certain
stable states (which turn out to be, in a sense, decay products
of the other states) will survive during the time evolution of
the system.

In quantum optics the quantum interference of Gaussian states in
phase space was investigated by Schleich and coworkers
\cite{sch1,sch2} and the decay of these quantum coherences was
studied in Refs. \cite{buz1,buz2} (for a recent review on
quantum interference in phase space see the work of Bu\v zek and
Knight \cite{buz3}).

The predictability sieve was recently proposed \cite{paz2} and
implemented for a harmonic oscillator with the resulting
evolution of the reduced density matrix 
\cite{HPZ,UnZ}. For a weakly
damped harmonic oscillator, pure states selected by the
predictability sieve, called preferred states, turn out to be the coherent 
states
\cite{paz2,Zur}.
Decoherence is caused by the loss of
phase coherence between the preferred quantum states in the Hilbert
space
of the system due to the interaction with the environment
\cite{Zur2,JoZ}. 
Preferred states are singled out by their stability (measured, for example, by
the rate of predictability loss -- the rate of entropy increase) under the
joint influence of the environment and the
self-Hamiltonian \cite{Zur}. Thus, the strength and nature of
the coupling with the environment play a crucial role in selecting
preferred states, which -- given the distance-dependent nature of
typical interactions -- explains the special function of the position
observable \cite{Zur2,Zur}. Coupling with the environment
also sets the decoherence time scale -- the
time on which quantum interference between preferred
states disappears \cite{Zur2,JoZ,paz1,paz2,Zur,zur1}.
Classicality is then an emergent property of an open quantum
system. 

The interaction with the environment also induces fluctuations
in the evolution of the system. There is therefore
a certain degree of tension between the demands of decoherence
and approximate classical predictability: decoherence requires
interaction with an environment, which inevitably produces
fluctuations, but classical predictability requires that these
fluctuations be small. 
The von Neumann entropy is an
ideal parameter with which to characterize the decoherence, 
i. e. the rapid decay of off-diagonal coherences. 
In \cite{AnH} an information-theoretic
measure of the size of these fluctuations was proposed -- the
Shannon information of the $Q_H$ distribution of the density
matrix $\hat\rho $ for the distinguished system (see Sec. 4). 
This measure of uncertainty is in fact bounded from below by the
von Neumann entropy of $\hat\rho .$ This suggests that the von
Neumann entropy is the key to understanding the connection
between decoherence and fluctuations: it limits the amount of
decoherence from above but bounds the size of the fluctuations
from below. Large entropy, which is a signal of destruction of
interference, therefore permits good decoherence,
but leads to large fluctuations; on the other hand, small
entropy allows small fluctuations, but the amount of decoherence
is also small. Environmentally induced fluctuations are
inescapable if one is to have decoherence.
Although there is some tension, there is a broad compromise
regime in which decoherence and classical predictability can
each hold extremelly well \cite{AnH}.

As we have already mentioned, besides the von Neumann entropy $S$ 
(\ref{neuent1}), there is another quantity
which can measure the degree of mixing or purity of quantum states. It
is the linear entropy $S_l$ (\ref{linent}).
For pure states $S_l=0$ and for a statistical mixture $S_l>0.$ 
The increasing of the linear entropy $S_l$
(as well as of the von Neumann entropy $S$) due to the interaction with the
environment is associated with the
decoherence phenomenon, given by the diffusion
process \cite{paz1,paz2}.
Due to the decoherence, the macroscopic systems obey
essentially classical equations of motion, despite the quantum
mechanical nature of the underlying microscopic dynamics.  
Dissipation increases the entropy and the pure states are converted into
mixed states.
The rate of entropy production is given by
\bea {\dot S}_l(\hat\rho)=-2{\rm Tr}(\hat\rho\dot{\hat \rho})=-2{\rm Tr}
(\hat\rho L(\hat\rho)), \label{entpro1}\eea
where $L$ is the evolution operator.
According to Zurek's theory, the maximally predictive states are the pure
states which minimize the entropy production in time.
These states remain least affected by the openness of the system 
and form the preferred set of
states in the Hilbert space of the system, known as the
pointer basis. Their evolution is predictible with the principle of least
possible entropy production.

Using Eq. (\ref{mast}), in our model the rate of entropy production 
(\ref{entpro1}) is given by:
\bea \dot S_l(t)={4\over\hbar^2}
[D_{pp}{\rm Tr}(\hat\rho^2\hat q^2-\hat\rho\hat q\hat\rho\hat q)
~~~~~~~~~~~~~~~~~~~~\nonumber \\
+D_{qq}{\rm Tr}(\hat\rho^2\hat p^2-\hat\rho\hat p\hat\rho\hat p) -
D_{pq}{\rm Tr}(\hat\rho^2(\hat q\hat p+\hat p\hat q)-
2\hat\rho\hat q\hat\rho\hat p) -
{\hbar^2\lambda\over 2}{\rm Tr}(\hat\rho^2)]. \label{entpro3} \eea
When the state remains approximately pure $(\hat\rho^2\approx\hat\rho),$ 
we obtain:
\bea \dot S_l(t)={4\over\hbar^2}(D_{pp}\sigma_{qq}(t)+D_{qq}\sigma_{pp}(t)-
2D_{pq}\sigma_{pq}(t)-{\hbar^2\lambda\over 2})\ge 0,\label{entpro4}\eea 
according to (\ref{has3}). 
We see that $\dot S_l(t)=0$ when the condition (\ref{has2}) of 
purity
for any time $t$ is fulfilled. The entropy production $S_l$ is  
also equal to 0 at $t=0$ if the initial state is a pure state. We have 
shown in the preceding section that the only initial states which remain pure 
for any $t$ are 
the correlated coherent states (squeezed coherent states) and therefore, we 
can state that in the Lindblad
theory for the open quantum harmonic oscillator the correlated 
coherent states, which are generalized coherent states, are the maximally 
predictive states. Our result generalizes 
the previous results which assert that for 
many models of quantum Brownian motion in the high 
temperature limit the usual coherent states correspond to minimal entropy
production and, therefore, they are the maximally predictive states.
In our model the coherent states can be obtained
as a particular case of the correlated coherent states by taking
$\mu=0,$ so that the correlation coefficient $r=0$ (see Eq. (\ref{corcoe})).
 
Paz, Habib and Zurek \cite{paz1,paz2} 
considered the harmonic oscillator undergoing 
quantum Brownian motion in the Caldeira-Leggett model
and concluded that the 
minimizing states which are the initial states generating the least amount 
of von Neumann or linear entropy and, therefore, the most predictible or 
stable ones under evolution in the presence 
of an environment are the ordinary coherent states. 
Using an information-theoretic measure of uncertainty for quantum systems,
Anderson and Halliwell showed in \cite{AndH} that the minimizing states
are more general Gaussian states. Anastopoulos and Halliwell 
\cite{AnH} offered an alternative characterization of these states by noting
that these states 
minimize the generalized uncertainty relation.
According to this assertion, we can say that in the Lindblad model the 
correlated coherent states (squeezed coherent states) are the most stable 
states which minimize the 
generalized uncertainty relation (\ref{genun}). Our result 
confirms that one of 
\cite{AnH}, 
but the model used in \cite{AnH} is different, namely 
the open quantum system is a particle
moving in a harmonic oscillator potential and linearly coupled to an
environment consisting of a bath of harmonic oscillators in a thermal 
state. 
At the same time 
we remind that the Caldeira-Leggett model considered in \cite{paz1,paz2}
violates the positivity of the density operator at short time scales
\cite{Amb,dio}, whereas
in the Lindblad model considered here the property of positivity is
automatically fulfilled. 

The rate of predictability loss, measured by the rate of linear entropy 
increase, for a damped harmonic oscillator is also calculated in the framework
of Lindblad theory in Ref. \cite{par}. The initial states which minimize
the predictability loss are identified as quasi-free states with a symmetry
dictated by the environment diffusion coefficients. For an isotropic 
diffusion in phase space, the coherent states or mixtures of coherent states
are selected as the most stable ones.  

In order to generalize the results of Zurek and collaborators, the entropy 
production
was considered by Gallis \cite{gal} within the Lindblad theory of open quantum
systems, treating environment effects perturbatively. Gallis considered the
particular case with $D_{pq}=0$ and found out that the squeezed states 
emerge as the most stable states for intermediate times compared to the
dynamical time scales. The amount of squeezing decreases with time, so that
the coherent states are most stable for large time scales.
For $D_{pq}\not=0$ our results generalize the ones of Gallis and establish
that the correlated coherent states are the most stable under the evolution 
in the presence of an environment.  

\section{Summary and concluding remarks}

Recently there is a revival of interest in quantum Brownian motion as a
paradigm of quantum open systems. The possibility
of preparing systems in macroscopic quantum states led to the problems of
dissipation in tunneling and of loss of quantum coherence (decoherence). These 
problems are intimately related to the issue of quantum to classical 
transition and 
all of them point the necessity of a better understanding of open quantum 
systems. 
The Lindblad theory provides a selfconsistent 
treatment of damping as a general extension of quantum mechanics to open 
systems and gives the possibility to extend the model of quantum Brownian 
motion. In the present paper we have studied the one-dimensional harmonic 
oscillator with dissipation within the framework of this theory. 
From the master equation of the damped quantum oscillator we have
derived the corresponding Fokker-Planck equation in the Wigner $W$ 
representation. The obtained equation describes an Ornstein-Uhlenbeck 
process and the Wigner function is a two-dimensional Gaussian.
We have then obtained the explicit form of the density operator from the
master and Fokker-Planck equations. The density operator
in a Gaussian form is a function of the position and momentum operators
in addition to several time dependent
factors.
In the long time limit the density operator approaches a thermal state.
Then the density operator has been used
to calculate the von Neumann entropy and the effective temperature. 
The temporal behaviour of these quantities shows how they approach
their equilibrium values.
Following Anderson, Anastopoulos and Halliwell \cite{AndH,AnH}, we have 
put the von Neumann entropy in association
with uncertainty and linear entropy.
The Wehrl entropy generates an informatical-theoretical measure of the
size of the intrinsic state fluctuations. As coherent states are most
robust in dissipative environment \cite{paz1,paz2,Zur,AnH},
this suggests the utility of the Wehrl entropy in characterizing the
decoherence process in quantum mechanics.
We have also shown that the only states which stay pure during the evolution
in time of the system are the correlated coherent states
(squeezed coherent states) under the condition
of a special 
choice of the environment coefficients. These states are also connected with
the decoherence phenomenon and they are the most stable under the evolution
in the presence of the environment.
The obtained results in the framework of the Lindblad theory can
be used for the description in more details of the connection
between uncertainty, decoherence and correlations of open quantum systems 
with their environment.

{\bf Acknowledgements}

The author would like to thank 
J. J. Halliwell, A. Sandulescu, W. Scheid and H. Scutaru for fruitful 
discussions and the referees for suggestions and comments.


\newpage

\begin{thebibliography}{99}

\bibitem{Car} 
H. J. Carmichael, {\it An Open System Approach to Quantum Optics} 
(Springer,Verlag, 1993)
 
\bibitem{Zur2} 
W. Zurek, Phys. Rev. D {\bf 24}, 1516 (1981); {\bf 26}, 1862 
(1982)

\bibitem{GeH} 
M. Gell-Mann and J. B. Hartle, {\it Complexity, Entropy
and the Physics of Information, SFI Studies in the Sciences of Complexity},
Vol. VIII, ed. by W. Zurek (Addison Wesley, Reading, 1990); 
Proceedings of the Third International Symposium on the Foundations of
Quantum Mechanics in the Light of New Technology, ed. by S. Kobayashi, 
H. Ezawa,
Y. Murayama and S. Nomura (Physical Society of Japan, Tokyo, 1990);
Phys. Rev. D {\bf 47}, 3345 (1993)
 
\bibitem{JoZ} 
E. Joos and H. D. Zeh, Z. Phys. B {\bf 59}, 229 (1985)

\bibitem{Zeh} 
H. D. Zeh, Phys. Lett. A {\bf 172}, 189 (1993)

\bibitem{gisp}
N. Gisin and I. Percival, Phys. Lett. A {\bf 175}, 144 (1993)
 
\bibitem{omn}
R. Omn\` es, Phys. Rev. A {\bf 56}, 3383 (1997)

\bibitem{h}
R. W. Hasse, J. Math. Phys. {\bf 16}, 2005 (1975) 

\bibitem{d}
E. B. Davies, {\em Quantum Theory of Open Systems} (Academic Press, New 
York, 1976)

\bibitem{s}
H. Spohn, Rev. Mod. Phys. {\bf 52}, 569 (1980)

\bibitem{de}
H. Dekker, Phys. Rep. {\bf 80}, 1 (1981) 

\bibitem{l}
K. H. Li, Phys. Rep. {\bf 134}, 1 (1986) 

\bibitem{rev}
A. Isar, A. Sandulescu, H. Scutaru, E. Stefanescu and W. Scheid, Int. J. 
Mod. Phys. E {\bf 3}, 635 (1994) 

\bibitem{d1}
H. Dekker, Physica A {\bf 95}, 311 (1979)

\bibitem{l1}
G. Lindblad, Commun. Math. Phys. {\bf 48}, 119 (1976) 

\bibitem{HPZ} B. L. Hu, J. P. Paz and Y. Zhang, Phys. Rev. D {\bf 45}, 2843
(1992); {\bf 47}, 1576 (1993)

\bibitem{paz1}
J. P. Paz, S. Habib and W. Zurek, Phys. Rev. D {\bf 47}, 488 (1993)  

\bibitem{PaZ} J. P. Paz and W. Zurek, Phys. Rev D {\bf 48}, 2728 (1993)

\bibitem{UnZ} W. G. Unruh and W. Zurek, Phys. Rev. D {\bf 40}, 1071 (1989)
 
\bibitem{paz2}
W. Zurek, S. Habib and J. P. Paz, Phys. Rev. Lett. {\bf 70}, 1187 (1993)

\bibitem{Zur} 
W. Zurek, 
Phys. Today {\bf 44}, No. 10, 36 (1991); {\bf 46}, No. 12, 81 (1993);
Prog. Theor. Phys. {\bf 89},
281 (1993); {\it Physical Origins of Time Asymmetry},
ed. by J. Halliwell, J. Perez-Mercader and W. Zurek (Cambridge
University Press, Cambridge, 1994)

\bibitem{Gri} 
R. Griffiths, J. Stat. Phys. {\bf 36}, 219 (1984)

\bibitem{Omn} 
R. Omn\`es, {\it The Interpretation of Quantum Mechanics}
(Princeton University Press, Princeton, 1994);
Rev. Mod. Phys. {\bf 64}, 339 (1992) 
 
 \bibitem{DoH} 
H. F. Dowker and J. J. Halliwell, Phys. Rev. D {\bf 46}, 1580
(1992)
 
\bibitem{l2}
G. Lindblad, Rep. Math. Phys. {\bf 10}, 393 (1976)

\bibitem{ss}
A. Sandulescu and H. Scutaru, Ann. Phys. (N. Y.) {\bf 173}, 277 (1987)

\bibitem{i3}
A. Isar, A. Sandulescu and W. Scheid, J. Math. Phys. {\bf 34}, 3887 
(1993) 

\bibitem{i1}
A. Isar, A. Sandulescu and W. Scheid, J. Phys. G: Nucl. Part. Phys. 
{\bf 17}, 385 (1991) 

\bibitem{g}
M. R. Gallis, Phys. Rev. A {\bf 48}, 1028 (1993) 

\bibitem{i2}
A. Isar, W. Scheid and A. Sandulescu, J. Math. Phys. {\bf 32}, 2128 
(1991) 

\bibitem{wig}
A. Isar, Helv. Phys. Acta {\bf 67}, 436 (1994)

\bibitem{ent}
A. Isar, Helv. Phys. Acta {\bf 68}, 225 (1995)

\bibitem{vlas}
A. Isar, A. Sandulescu and W. Scheid, Int. J. Mod. Phys. B {\bf 10}, 2767 
(1996) 

\bibitem{h1}
H. Haken, {\em Handbuch der Physik,} vol.XXV/2C (Springer,
Berlin, 1970)  

\bibitem{lo}
W. H. Louisell, {\em Quantum Statistical Properties of Radiation} (Wiley, 
New York, 1973)

\bibitem{gr}
R. Graham, {\em Springer Tracts in Mod. Phys.} {\bf 66} (Springer, Berlin, 
1973) p. 1
  
\bibitem{h2}
F. Haake, {\em Springer Tracts in Mod. Phys.} {\bf 66} (Springer, Berlin, 
1973) p. 98 

\bibitem{ga}
C. W. Gardiner, {\em Quantum Noise} (Springer, Berlin, 1991)

\bibitem{AndH} 
A. Anderson and J. J. Halliwell, Phys. Rev. D {\bf 48}, 2753 (1993)
 
\bibitem{AnH} 
C. Anastopoulos and J. J. Halliwell, Phys. Rev. D {\bf 51}, 6870 (1995)

\bibitem{zel}
B. Ya. Zel'dovich, A. M. Perelomov and V. S. Popov, Zh. Eksp. Teor. Fiz.
{\bf 55}, 589 (1968) (Engl. transl. Sov. Phys. -- JETP {\bf 28},
308 (1969))

\bibitem{hug} 
P. Huguenin, Helv. Phys. Acta {\bf 51}, 346 (1978) 

\bibitem{halzou}
J. J. Halliwell and A. Zoupas, Phys. Rev. D {\bf 52}, 7294 (1995) 

\bibitem{gal} 
M. R. Gallis, Phys. Rev. A. {\bf 53}, 655 (1996) 

\bibitem{haa}
F. Haake and R. Reinbold, Phys. Rev. A {\bf 32}, 2462 (1985)

\bibitem{tal}
P. Talkner, Ann. Phys. (N.Y.) {\bf 167}, 39 (1986)

\bibitem{a2}
G. S. Agarwal, Phys. Rev. A {\bf 4}, 739 (1971)

\bibitem{da}
S. Dattagupta, Phys. Rev. A {\bf 30}, 1525 (1984) 

\bibitem{lz}
N. Lu, S. Y. Zhu and G. S. Agarwal, Phys. Rev. A {\bf 40}, 258 (1989) 

\bibitem{j}
S. Jang, Nucl. Phys. A {\bf 499}, 250 (1989) 

\bibitem{CaL} 
A. O. Caldeira and A. J. Leggett,  Physica A {\bf 121}, 587 (1983)
 
\bibitem{Dio7} 
L. Di\'osi, Europhys. Lett. {\bf 22}, 1 (1993); Physica A {\bf 199}, 517 (1993)

\bibitem{Amb} 
V. Ambegaokar, Ber. Bunsenges. Phys. Chem. {\bf 95}, 400 (1991)
 
\bibitem{w}
M. C. Wang and G. E. Uhlenbeck, Rev. Mod. Phys. {\bf 17}, 323 (1945)  

\bibitem{a1}
G. S. Agarwal, Phys. Rev. A {\bf 3}, 828 (1971) 

\bibitem{j1}
S. Jang, Physica A {\bf 175}, 420 (1991) 

\bibitem{lw}
W. H. Louisell and R. L. Walker, Phys. Rev. {\bf 137}, B 204 (1965) 

\bibitem{m}
J. E. Moyal, Proc. Cambridge Philos. Soc. {\bf 45}, 99 (1949) 

\bibitem{wx}
R. M. Wilcox, J. Math. Phys. {\bf 8}, 962 (1967) 

\bibitem{hw}
H. Weyl, {\it The Theory of Groups and Quantum Mechanics}
(Dover, New York, 1950)

\bibitem{clm}
C. L. Mehta, J. Math. Phys. {\bf 5}, 677 (1964) 

\bibitem{h5}
M. Hillery, R. F. O 'Connell, M. O. Scully and E. P. Wigner, 
Phys. Rep. {\bf 106}, 121 (1984)

\bibitem{rh}
H. S. Robertson and M. A. Huerto, Phys. Rev. Lett. {\bf 23}, 825 (1969) 

\bibitem{dodkur} 
V. V. Dodonov, E. V. Kurmyshev and V. I. Manko,
Phys. Lett. {\bf 79A}, 150 (1980)

\bibitem{raj} 
A. K. Rajagopal, Phys. Lett. A {\bf 228}, 66 (1997)

\bibitem{bia1} 
I. Bialynicki-Birula and J. Mycielski, Commun. Math. Phys. {\bf
44}, 129 (1975)

\bibitem{bia2} 
I. Bialynicki-Birula, Phys. Lett. A {\bf 103}, 253 (1984)

\bibitem{buz4} 
V. Bu\v zek, C. H. Keitel and P. L. Knight, Phys. Rev. A 
{\bf 51}, 2575, 2594 (1995)

\bibitem{orl} 
A. Orlowski, H. Paul and B. Boehmer, Opt. Commun. {\bf 138}, 311 (1997)

\bibitem{Hus} 
K. Husimi, Proc. Phys. Math. Soc. Japan {\bf 22}, 264 (1940)

\bibitem{Hal1} 
J. J. Halliwell, Phys. Rev. D {\bf 46}, 1610 (1992)

\bibitem{Weh} 
A. Wehrl, Rep. Math. Phys. {\bf 16}, 353 (1979)

\bibitem{Lie} 
E. H. Lieb, Commun. Math. Phys. {\bf 62}, 35 (1978)

\bibitem{Cov} 
T. M. Cover and J. A. Thomas,
{\it Elements of Information Theory} (Wiley, New York, 1991).

\bibitem{dodman1} 
V. V. Dodonov and V. I. Man'ko,
Proc. Lebedev Phys. Inst. of Sciences,
{\bf 183}, 5 (1987)

\bibitem{PaS}
S. Paraoanu and H. Scutaru, private communication

\bibitem{spo} 
H. Spohn, J. Math. Phys {\bf 19}, 1227 (1978)

\bibitem{schr}
E. Schr\"odinger, Ber. Kgl. Akad. Wiss. (Berlin, 1930) p. 296

\bibitem{rob}
H. P. Robertson, Phys. Rev {\bf 35}, 667A (1930); {\bf 46}, 794 (1934)

\bibitem{huds}
R. L. Hudson, Rep. Math. Phys. {\bf 6}, 249 (1974)

\bibitem{scut} 
H. Scutaru, Phys. Lett. A {\bf 141}, 223 (1989); {\bf 167}, 326 (1992) 

\bibitem{dek} 
H. Dekker, Phys. Lett. A {\bf 80}, 369 (1980) 

\bibitem{has1} 
R. W. Hasse, Nucl. Phys. A {\bf 318}, 480 (1979) 

\bibitem{has2} 
R. W. Hasse, Phys. Lett. B {\bf 85}, 197 (1979) 

\bibitem{deval} 
H. Dekker, M. C. Valsakumar, Phys. Lett. A {\bf 104}, 67 (1984)

\bibitem{sud}
E. C. G. Sudarshan, C. B. Chiu and G. Bhanathi, Phys. Rev. A
{\bf 52}, 43 (1995)

\bibitem{dodman} 
V. V. Dodonov and V. I. Man'ko, {\it Group Theory,
Gravitation and Elementary Particle Physics,} Proc. Lebedev Phys.
Inst. of Sciences {\bf 167},
ed. by A. A. Komar (Nova Science, Commack, New York, 1987) p. 7

\bibitem{sch1}
W. P. Schleich, D. F. Walls and J. A. Wheeler, Phys. Rev. A
{\bf 38}, 1177 (1988)

\bibitem{sch2}
W. P. Schleich, M. Pernigo and F. L. Kien, Phys. Rev. A
{\bf 44}, 2172 (1991)

\bibitem{buz1}
V. Bu\v zek, A. V. Barrenco and P. L. Knight, Phys. Rev. A
{\bf 45}, 6570 (1992)

\bibitem{buz2}
V. Bu\v zek, G. Adam and G. Drobn\'y, Phys. Rev. A
{\bf 54}, 804 (1996)

\bibitem{buz3}
V. Bu\v zek and P. K. Knight, Progress in Optics, Vol. XXXIV, ed. by
E. Wolf (Elsevier, Amsterdam, 1995) p. 1

\bibitem{zur1} 
W. H. Zurek, {\it Frontiers of Nonequilibrium Statistical
Mechanics}, ed. by G.
T. Moore and M. O. Scully (Plenum, New York, 1986)

\bibitem{dio}
L. Di\'osi, Physica A {\bf 199}, 517 (1993) 

\bibitem{par}
S. Paraoanu and H. Scutaru, Phys. Lett. A {\bf 238}, 219 (1998)

\end{thebibliography}
\end{document}